# Defying the Gibbs Phase Rule: Evidence for an Entropy-Driven Quintuple Point in Colloid-Polymer Mixtures

V. F. D. Peters,[1] M. Vis,[1,2] Á. González García,[1,3] H. H. Wensink,[4] and R. Tuinier[1,3,*]

[1]*Laboratory of Physical Chemistry, Department of Chemical Engineering and Chemistry*
*& Institute for Complex Molecular Systems, Eindhoven University of Technology,*
*P.O. Box 513, 5600 MB Eindhoven, Netherlands*
[2]*Laboratoire de Chimie, École Normale Supérieure de Lyon, 69364 Lyon CEDEX 07, France*
[3]*Van 't Hoff Laboratory for Physical and Colloid Chemistry, Department of Chemistry*
*& Debye Institute for Nanomaterials Science, Utrecht University, Padualaan 8, 3584 CH Utrecht, Netherlands*
[4]*Laboratoire de Physique des Solides—UMR 8502, CNRS & Université Paris-Saclay, 91405 Orsay, France*

Using a minimal algebraic model for the thermodynamics of binary rod-polymer mixtures, we provide evidence for a quintuple phase equilibrium; an observation that seems to be at odds with the Gibbs phase rule for two-component systems. Our model is based on equations of state for the relevant liquid crystal phases that are in quantitative agreement with computer simulations. We argue that the appearance of a quintuple equilibrium, involving an isotropic fluid, a nematic and smectic liquid crystal, and two solid phases, can be reconciled with a generalized Gibbs phase rule in which the two intrinsic length scales of the athermal colloid-polymer mixture act as additional field variables.

Dispersions of rodlike colloidal particles, be they of biological origin such as tobacco mosaic viruses [1] or filamentous bacteriophage fd viruses [2], or of synthetic nature, like cellulose nanocrystals [3], boehmite [4] or silica colloids [5], may form liquid crystal phases characterized by a concurrence of fluidity and long-range crystalline order [6–8]. The phase states of these dispersions are strongly influenced by colloid concentration and the rod length-to-diameter aspect ratio. In order to study the intricate interplay of various entropic effects, rod-shaped colloids are usually modeled as hard spherocylinders (HSCs) or infinitely thin needles only interacting through their excluded volume [6–14]. In real-world systems the phase behavior is however influenced by additional soft or long-range interactions, the presence of other species, or by size or shape dispersity. The recent decade has seen a quest for simple interparticle potentials that capture these effects and are capable of generating complex multiphase behavior and exotic crystalline structures [15,16]. Often these potentials are inspired by interactions between core-corona particles comprising multiple length sales [15,16]. These potentials have been shown to stabilize novel phase morphologies such as cluster crystals [17] and quasicrystals [18].

Here we explore such complexity by examining the effect of depletion interactions induced by the presence of non-adsorbing polymers [19,20], which are known to considerably enrich conventional colloidal phase behavior [21–27]. The depletion interaction is caused by a reduction of the number of possible configurations of the polymer chains within the so-called depletion zone around each colloidal particle. When the depletion zones between adjacent colloids overlap, the free volume available for the polymers is increased, and thus the polymer entropy increases as well, leading to an effective attraction between the colloids. The depletion interaction is often applied to modulate attractions with well-defined range and strength between colloidal particles [20]. Colloid-polymer mixtures are known to display a wealth of phase states and are widely used as the basis for directing mesoscale order and self-assembly in soft matter [28–30]. Moreover, solid nanoparticles mixed with polymers play a prominent role in various industrial formulations as well as in biological systems [31–33].

The influence of the depletion attraction on colloidal phase behavior can be predicted from free volume theory (FVT) [19,20], which accounts for the presence of polymers by correlating the mean "free" volume that is accessible to the polymer to colloidal structure. As this theory is perturbative in nature it requires equations of state of HSCs for all involved phases. For certain phase states there is however a lack of tractable analytical equations of state and therefore the rod-polymer phase behavior could only be theoretically scrutinized for a limited number of phases and rod-polymer size ratios [34–37]. Building on previous modeling efforts for pure rod dispersions [38] and disk-polymer mixtures [39], we develop a tractable quantitative theory to map out the complete phase diagram for two-component rod-polymer mixtures and provide evidence for stable quadruple and quintuple equilibria, that have hitherto gone unnoticed in previous modeling attempts [34–37]. The microscopic conditions (rod aspect ratio and rod-polymer size ratio), under which these multiphase equilibria are found, are within reach for experimentally relevant rod-polymer mixtures.

First we briefly discuss the phase behavior found in pure rod dispersions and the equations of state used to describe the different phases. Then we explain how phase coexistences are predicted in the colloid-polymer mixtures.

Colloidal rods are represented as hard spherocylinders with length $L$ and diameter $D$ (total length, including hemispherical ends, equals $L + D$). From simulation results of HSC suspensions it is known that, with increasing concentration, isotropic (I), nematic (N), smectic A (SmA), AAA crystal, and ABC crystal phases may be observed, as indicated in the top panel of Fig. 1 [7,13,14]. In the I phase the rods have random orientations, while in the other phases they are aligned along a common nematic director. Both I and N phases are fluids and do not possess long-range positional order. Approximate equations of state for these states are available from both scaled particle theory (SPT) [9] and Parsons-Lee (PL) theory [10–12]. Using a Gaussian approximation for the distribution of rod angles with respect to the nematic director, one can construct algebraic equations of state [37]. Given that PL theory tends to be more accurate for weakly anisotropic rods while SPT has proven more reliable for long rods, we employ a straightforward sigmoidal interpolation procedure in order to accurately cover the full range of aspect ratios [40]. We emphasize that the observed phase behavior is robust and is not qualitatively affected by details of the rod-polymer model (see Fig. S1 in the Supplemental Material [41]).

In the SmA, AAA, and ABC phase the particles are confined in layers and the nematic director is perpendicular to the layers. For the SmA phase there is no long-range positional order within the layers, while in the AAA and ABC phases the particles are ordered hexagonally. In the AAA phase the rods of adjacent layers are stacked on top of each other, while in the ABC phase they are stacked in between the rods of adjacent layers. For all these three phases we have used an extended cell theory similar to Graf and Löwen [38] and the results were subsequently cast into algebraic equations [40]. For the SmA phase this includes a thermodynamic description of 2D disks that captures the in-plane fluidity of rods projected onto the smectic plane. In the sphere limit of $L/D \to 0$ our equations of state for the isotropic and ABC phases become equivalent to those of, respectively, a hard sphere fluid and a fcc crystal.

Phase coexistence between two phases can be established by imposing mechanical and chemical equilibrium expressed by equality of osmotic pressure $\Pi$ and chemical potential $\mu$. In Fig. 1 we show the predictions for the binodals (solid curves) obtained from the analytical equations of state as a function of volume fraction $\eta$ and aspect ratio $D/L$. Comparing predictions with computer simulation results of Bolhuis and Frenkel (data points) [14] we find excellent agreement. We remark that our model asserts that all phase transitions are first order. While this is true for a large section of the phase diagram, there is still some discussion as to the nature of the N-SmA transition at small $D/L$. Density functional theory predictions suggest that it is second order below a certain critical point [42–45], while simulation results are inconclusive on the transition order [14].

In the colloid-polymer mixtures the nonadsorbing polymers are modeled as penetrable hard spheres with radius $\delta$ [20]. The polymers behave as hard spheres with respect to the colloids, yet freely overlap each other. Since all particle interactions are of the excluded-volume type, all phase transitions are driven by entropy alone so that the mixture is strictly athermal ($\Delta H = 0$, with $H$ the enthalpy of the mixture). Previous calculations on colloid-polymer models that include explicit polymer-polymer interactions have demonstrated that the penetrable hard sphere approximation for the polymer works well for the relatively small polymers we consider here [20,46,47]. In accordance with FVT, we use a semigrand canonical ensemble, where the colloid-polymer mixture is held in contact with a polymer solution reservoir through a semipermeable membrane that is impermeable to the rods [19]. FVT states

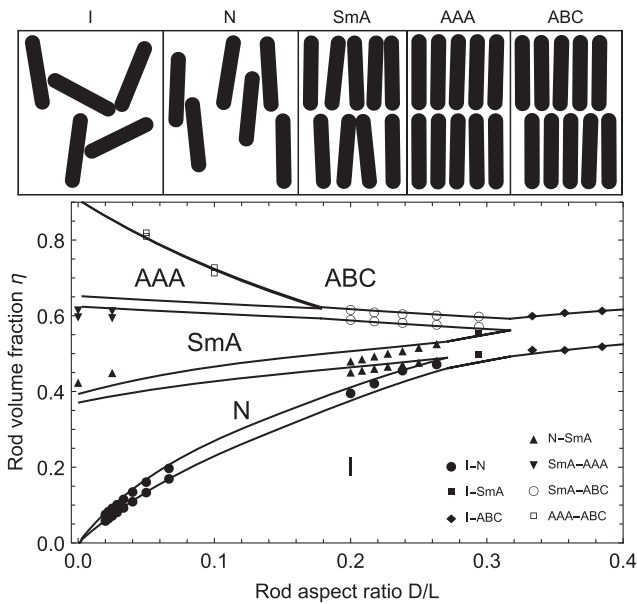

FIG. 1. Phase behavior of hard spherocylinders as a function of rod volume fraction $\eta$ and aspect ratio $D/L$ from both theory (solid curves) [40] and simulation (data points) [14]. The stable phases include the isotropic (I), nematic (N), smectic A (SmA), AAA crystal, and ABC crystal phase.

that the semigrand potential $\Omega$ of the mixture can be approximated as follows [19]:

$$\omega = \frac{\Omega v_c}{k_B T V} = f^0 - \alpha \tilde{\Pi}^R, \quad (1)$$

where $\omega$ is the normalized semigrand potential, $v_c$ the rod volume, $k_B T$ the thermal energy, $V$ the volume of the system, $f^0$ the normalized Helmholtz free energy for a pure colloid dispersion, $\alpha = \langle V_{\text{free}} \rangle / V$ the average fraction of the system volume available to polymers, and $\tilde{\Pi}^R$ the normalized osmotic pressure of the polymers in the reservoir, which is proportional to the reservoir polymer volume fraction $\phi^R$ via $\tilde{\Pi}^R = \phi^R (3\Gamma - 1)/(2q^3)$. Here we have defined a normalized total rod length $\Gamma$ and polymer size $q$ relative to the rod diameter: $\Gamma = 1 + L/D$ and $q = 2\delta/D$. The rod aspect ratio and the colloid-polymer size ratio are the two basic length scales of our model and the phase behavior depends sensitively on particle geometry through variation of $\Gamma$ and $q$. We estimate $\alpha$ by assuming that the average free volume for the polymer in the system is unaffected by the presence of the polymers. From scaled particle theory we can express the free volume fraction as [34]

$$\alpha = (1 - \eta) \exp[-Q], \quad (2)$$

where $\eta$ is the colloid volume fraction and $Q = 3yq(2\Gamma + q\Gamma + q)/(3\Gamma - 1) + 18y^2 q^2 \Gamma^2/(3\Gamma - 1)^2 + 2q^3 \tilde{\Pi}^0/(3\Gamma - 1)$ with $y = \eta/(1 - \eta)$ and the superscript $^0$ again referring to the pure rod dispersion for the same phase type.

A representative set of phase diagrams for colloid-polymer mixtures is shown in Fig. 2 in terms of the polymer reservoir volume fraction $\phi^R$ versus the rod volume fraction $\eta$. The rod aspect ratio is fixed at $L/D = 12$ and the polymer-rod size ratio is set to $q = 0.4, 0.525$, and $0.57$. In the plots we show the binodals (solid curves) and three- and four-phase coexistences (dashed lines). In most cases the miscibility gaps widen as the polymer concentration is increased. At the points where two binodals coincide, we find three-phase coexistence. For instance at $q = 0.4$ (left panel), the miscibility gap of N-SmA and SmA-AAA coexistence widens as $\phi^R$ increases. At around $\eta = 0.4$–$0.65$ and $\phi^R \approx 0.05$ the binodals coincide and a triple N-SmA-AAA equilibrium emerges.

Increasing the polymer size brings about considerable qualitative changes in the phase diagram. For example, at $q = 0.57$ (right panel) the N-SmA binodal coincides with the I-N binodal leading to a triple I-N-SmA coexistence instead. Similarly the N-AAA and I-N-AAA coexistences are only present at the smaller $q = 0.4$, while the I-SmA binodals and I-SmA-AAA triphasic coexistence are only stable at $q = 0.57$. The intermediate polymer size of $q = 0.525$ (middle panel) marks the exact size ratio where all three binodals coincide at the same polymer reservoir concentration. This leads to an I-N-SmA-AAA four-phase coexistence that is reminiscent of the predictions for the disk-polymer systems [39].

The four-phase coexistence prompts us to contemplate the Gibbs phase rule for an athermal system, which reads

$$F = C - N + 1, \quad (3)$$

where $F$ is the number of degrees of freedom, $C$ is the number of different components, and $N$ is the number of coexisting phases. If we consider the solvent as an irrelevant background medium, we find that an effective two-component system ($C = 2$) should not be able to display a four-phase equilibrium, since this would imply $F$ being negative. Vega and Monson [48] proposed that for mixtures with anisotropic particles the relevant shape aspect ratios should be added in the Gibbs phase rule as an extra field variable. This has been used to explain the presence of triple points in pure dispersions of hard anisotropic particles. Using Monte Carlo simulations Akahane *et al.* [49] have demonstrated the validity of a generalized Gibbs phase rule by revealing a four-phase coexistence in a (thermal) one-component system with an anisotropic interaction potential. Inspired by these observations we propose the following generalized Gibbs phase rule for athermal systems:

$$F = C - N + S + 1, \quad (4)$$

where $S$ is the number of independent microscopic length scales influencing the free energy (see Supplemental Material [41] for a detailed derivation). In our particular rod-polymer model both the rod aspect ratio $L/D$ and colloid-polymer size ratio $q$ affect the anisotropic interaction between rods and must therefore be considered as additional "geometric" field variables, so that $S = 2$. Subsequently, the generalized Gibbs rule Eq. (4) implies that the maximum number of possible coexisting phases (corresponding to $F = 0$) should be five and that the quadruple equilibria still possess a single degree of freedom.

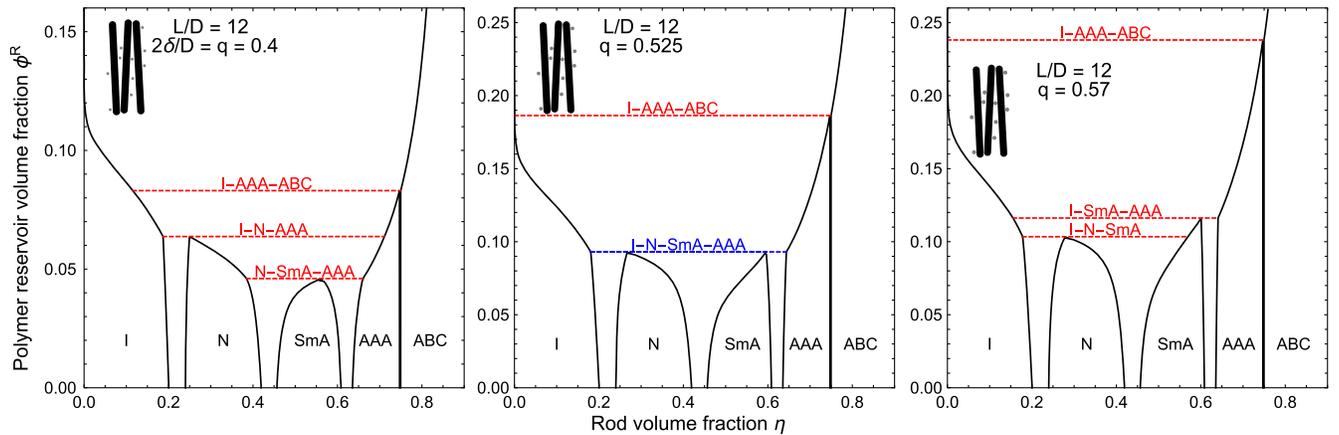

FIG. 2. Phase diagrams of colloid-polymer mixtures in terms of the colloid volume fraction $\eta$ and polymer reservoir volume fraction $\phi^R$ for colloidal rods of aspect ratio $L/D = 12$ and polymers of size $q = 2\delta/D = 0.4, 0.525$, and $0.57$. Binodals are displayed as solid curves, while three- and four-phase coexistences are indicated as dashed lines.

While Fig. 2 only shows the I-N-SmA-AAA coexistence at a particular $L/D$ and $q$, it is actually found for a range of values given that $F = 1$. This implies that there may be a single point in the parameter space where a *five*-phase coexistence is generated. Indeed, for $L/D = 6.086$ and $q = 0.470$ the I-N, N-SmA, SmA-AAA, and AAA-ABC binodals all coincide at the same polymer reservoir concentration, thus generating an I-N-SmA-AAA-ABC quintuple point as indicated in Fig. 3 (left graph). Even though the generalized Gibbs phase rule enables, in principle, the appearance of a quintuple point, it is not guaranteed to emerge for any system with $C = 2$ and $S = 2$, e.g., for the disk-polymer systems considered previously [39] a quintuple point was not observed.

While quintuple coexistence occurs at a uniform reservoir polymer concentration, the polymer contents in the individual coexisting phases, given by $\phi^S = \alpha \phi^R$, are distinctly different in view of their unequal free-volume fractions as shown in the right plot of Fig. 3.

It is however not possible to determine the relative volume fraction occupied by each of the coexisting phases with the conventional lever rule as this only applies for up to three phases in a two-component system. Given that the additional geometric variables are both degenerate across the phases, no new boundary conditions emerge to devise a generalized lever rule. This is similar to predicted triple points in pure dispersions of anisotropic particles. Instead there is a range of possible volume ratios depending on the total concentration of the particles. This includes the case where the volume of a phase vanishes. For instance in a system of pure HSCs at the I-N-SmA triple point and a total concentration near the lower limit of this coexistence, it is equally favorable to have an I-N, I-SmA, or I-N-SmA coexistence where the N and SmA phase will occupy a relatively small volume. For the similar triple point of thermal one-component systems it was argued how the exact phase volumes are determined from an additional boundary condition in the specific entropies [50]. Although there is no direct evidence of single-component athermal triple equilibria, the existence of such equilibria is hinted at by computer simulations for hard dumbbells and spherocylinders [48].

In conclusion, we have demonstrated the possibility of quintuple phase equilibria using a minimal model for an athermal rod-polymer mixture; an observation that is surprising in view of the Gibbs phase rule, that states that only two- and three-phase equilibria should be expected. We rationalize the observed quadruple and quintuple equilibria by contemplating a modified phase rule which includes the two intrinsic length scales of the mixture (rod aspect ratio and colloid-polymer size ratio) as additional field variables. A generalized lever rule, however, remains thus far elusive, and future efforts should be directed towards determining the precise phase fractions of the coexisting isotropic, nematic, smectic A, and ABC and AAA solid phase states. The $L/D$- and $q$-values at which the quintuple equilibrium is observed are, however, entirely realistic and should be realizable experimentally in colloidal rods mixed with nonadsorbing polymer. Strong size dispersity of the colloidal particles may suppress smectic and crystalline order in favor of the columnar or smectic B phase [2,5] or complicate the phase behavior through numerous multiphase coexistences, unseen in pure-component colloidal systems [51]. Our study further demonstrates that the introduction of nonadsorbing depletants may play a key role in stabilizing targeted crystal morphologies formed by anisotropic colloids [52,53] by rendering those high-density phases in simultaneous coexistence with low-density fluids thus facilitating their emergence and detection.

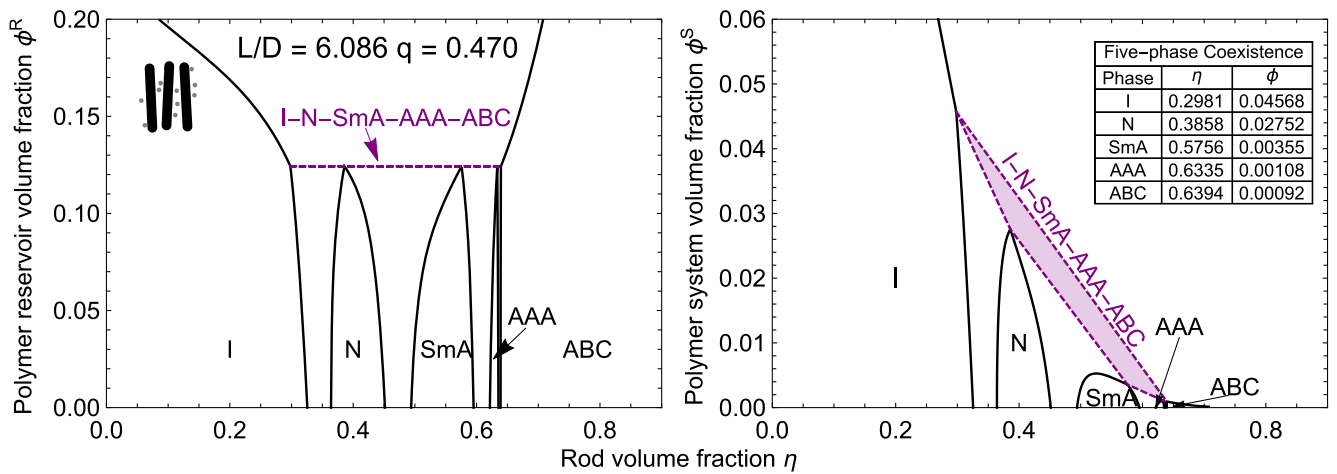

FIG. 3. Phase diagrams of rod-polymer mixtures in terms of the rod volume fraction $\eta$ and polymer reservoir (left panel) or system volume fraction (right panel), $\phi^R$ or $\phi^S$, for rod aspect ratio $L/D = 6.086$ and polymer size $q = 0.470$. Binodals are displayed as solid curves, while the five-phase coexistence is indicated by a dashed line. In the system representation (right panel) five-phase coexistence is predicted over the entire region enclosed by the dashed lines.

We thank H. N. W. Lekkerkerker for useful discussions and suggestions. M. V. acknowledges the Netherlands Organization for Scientific Research (NWO) for a Veni Grant (No. 722.017.005).